\documentclass[letter]{jpsj2}
\usepackage{graphicx}

\def\runauthor{Yunori {\sc Nisikawa}, Hiroaki {\sc Ikeda}
 and Kosaku {\sc Yamada}}

\title
{Possible Pairing Symmetry of Superconductor Na$_x$CoO$_2$$\cdot y$H$_2$O}

\author
{ 
Yunori  {\sc Nisikawa}\footnote{E-mail
address:nisikawa@spring8.or.jp}$^1$, Hiroaki {\sc Ikeda}$^2$ and Kosaku {\sc Yamada}$^2$
}

\inst
{$^1$Synchrotron Radiation Research Center, Japan Atomic Energy Research Institute, Mikazuki, Sayo, Hyogo 679-5148\\
$^2$Department of Physics, Kyoto University, Kyoto 606-8502}

\recdate
{
today
}
\abst{
To discuss the possibility that 
the superconductivities in Na$_x$CoO$_2$$\cdot y$H$_2$O 
are induced by electron correlation, 
we investigate the possible pairing symmetry 
based on the single-band Hubbard model whose 
dispersion of the bare energy band 
is obtained  using 
the band structure calculation of 
Na$_x$CoO$_2$$\cdot y$H$_2$O.
The superconducting transition temperature 
is estimated by solving the
${\rm\acute{E}}$liashberg equation.
In this equation, 
both normal and anomalous self-energies are calculated up to
the third-order terms with respect to the Coulomb repulsion. 
In the case of spin-singlet pairing, 
the candidate of pairing symmetry (the maximum eigen value 
$\lambda_{\rm max}^{\rm SS}$ of 
${\rm\acute{E}}$liashberg's equation) belongs to the $d$-wave(E$_2$ 
representation of D$_6$ group).
In the case of spin-triplet pairing, 
the candidate of pairing symmetry (the maximum eigen value 
$\lambda_{\rm max}^{\rm ST}$ of ${\rm\acute{E}}$liashberg's equation) 
belongs to the $f_{y(y^{2}-3x^{2})}$-wave (B$_1$ representation of D$_6$ group).
It is found that 
$\lambda_{\rm max}^{\rm SS}\simeq\lambda_{\rm max}^{\rm ST}$ 
and the transition temperatures of unconventional pairing state 
are estimated to be low
compared with the observed temperature within our simple model.

}
\kword{Na$_x$CoO$_2$$\cdot y$H$_2$O, superconductivity, vertex correction}

\begin{document}
\sloppy
\maketitle
Recently, Tanaka {\it et. al} discovered 
superconductivity in  Na$_x$CoO$_2$$\cdot y$H$_2$O
($x\simeq 0.35, y\simeq 1.4$) 
 with the transition temperature $T_{\rm c}\simeq 5$ K~\cite{rf:Taka}.
Na$_x$CoO$_2$$\cdot y$H$_2$O consists of a two-dimensional triangular 
lattice of cobalt ions formed by a network of edge-sharing CoO$_6$
octahedra, separated by layers of Na ions and H$_2$O molecules.
The superconductivity in this compound has been intensively
investigated~\cite{rf:z1,rf:z2,rf:z3,rf:z4,rf:z5,rf:z6,rf:z7,rf:z8,rf:z9,rf:z10,rf:heat,rf:nmr1,rf:nmr2,rf:nmr3,rf:ishi,rf:INY,rf:KTA,rf:Da}.
The normalized electronic specific heat data in the superconducting
state well fit the $T^{3}$ dependence~\cite{rf:heat}.
The NMR/NQR measurements 
have been performed by several groups~\cite{rf:nmr1,rf:nmr2,rf:nmr3}.
The conclusions of the groups have been contradictory 
regarding the pairing symmetry of the superconductivity, to date.
Fujimoto {\it et al.} observed an absence of the coherence peak just 
below $T_{\rm c}$~\cite{rf:nmr3} and quite recently, their observation 
has been confirmed by Ishida {\it et al.}~\cite{rf:ishi}
These results may suggest the possibility 
of unconventional superconductivity in Na$_x$CoO$_2$$\cdot y$H$_2$O. 
In this paper, we discuss the possibility that
the superconductivities in Na$_x$CoO$_2$$\cdot y$H$_2$O
are induced by electron correlation.
In our previous study~\cite{rf:INY}, we investigated
the superconducting instabilities of a two-dimensional triangular lattice 
in the Hubbard model with the next nearest 
hopping integral, by the perturbation 
theory with respect to Coulomb repulsion $U$ and 
fluctuation-exchange approximation(FLEX). 
We have found that  $f$-wave spin-triplet pairing is the most 
stable in wide ranges of next nearest neighbor hopping 
integrals and electron number densities.
The perturbation approach is sensitive to 
the dispersion of the bare energy band, by its nature.
This implies that the lattice structures and the band
filling play essential roles in the calculation of $T_{{\rm c}}$, and 
it is important to evaluate $T_{{\rm c}}$
on the basis of the detailed electronic structure in each system.
Therefore, we construct a two-dimensional Hubbard model whose  
dispersion of the bare energy band is obtained using the 
band structure calculation of Na$_x$CoO$_2$$\cdot y$H$_2$O.
Based on this Hubbard model, we estimate $T_{\rm c}$
by the perturbation theory with respect to Coulomb repulsion $U$.

Here, we briefly mention the band structure 
calculation of Na$_x$CoO$_2$$\cdot y$H$_2$O 
performed by Nisikawa {\it et al.}~\cite{rf:UsuNisi}
They used the full-potential linearlized 
augmented plane wave (FLAPW) method in the local density approximation(LDA).
In their calculation, the H$_2$O molecules were treated as mere spacers
 which realistically separate two CoO$_2$ layers in a hydrated compound and 
the Na ions were treated using the virtual crystal approximation.
The obtained band structure has two-dimensional properties.
The Fermi surfaces consist of a large ``troche''-like 
hole Fermi surface around the $\Gamma$ 
point and small cylindrical hole Fermi surfaces near the K points.
The Fermi level is very close to the van Hove singularity.
Co $t_{2g}$ orbitals
take the main part of the density of states near 
the Fermi energy.
The obtained Fermi surface is similar to the Fermi surface
of nonhydrated Na$_{0.5}$CoO$_2$ obtained by Singh~\cite{rf:Shin}.
But the volume of 
our cylindrical hole Fermi surface near the K point is larger than the
volume of Singh's one.
There is no experimental measurements
 about Fermi surface of Na$_x$CoO$_2$$\cdot y$H$_2$O to date.
Fermi surface is one of the important matters
to consider in investigating the mechanism of superconductivity.
So comparison of band structure calculation and experiments is desired.

We start from quasi-particle state and use the single-band Hubbard model 
for discussing the mechanism of superconductivity.
We obtain the dispersion $\epsilon({\bf k})$ of the bare energy band
by using the band structure calculation of
Na$_x$CoO$_2$$\cdot y$H$_2$O($x=0.35$).
$\epsilon({\bf k})$ can be expanded as 
$\epsilon({\bf k})=
-\sum_{{\bf R}\in L}t_{\bf R}\exp(i{\bf R}\cdot{\bf k})$,
where $L$ is the set of lattice vectors of a two-dimensional triangular
lattice.
We rescale length, energy, temperature and time by 
$a, t, \frac{t}{k_{\rm B}}, \frac{\hbar}{t}$ respectively
(where $a, t, k_{\rm B}, \hbar$ are the  lattice constants of the hexagonal
Co plane, third hopping integral, Boltzmann constant and 
Planck constant divided by $2\pi$, respectively).
Our Hamiltonian is written as
\begin{displaymath}
H=\sum_{{\bf k},\sigma}\left(\epsilon({\bf k})-\mu\right)
a_{{\bf k}\sigma}^{\dagger}a_{{\bf k}\sigma}+
\frac{U}{2N}
\sum_{\sigma\neq\sigma^{\prime}}
\sum_{{\bf k}_{1}{\bf k}_{2}{\bf k}_{3}{\bf k}_{4}}
\delta_{{\bf k}_{1}+{\bf k}_{2},
{\bf k}_{3}+{\bf k}_{4}}
a_{{\bf k}_{1}\sigma}^{\dagger}
a_{{\bf k}_{2}\sigma^{\prime}}^{\dagger}
a_{{\bf k}_{3}\sigma^{\prime}}a_{{\bf k}_{4}\sigma},
\end{displaymath}
where $\mu$ and $U$ are
 the chemical potential and the Coulomb repulsion, respectively.
The coefficients $t_{\bf R}$ are shown in Fig.~\ref{fig:exp}.
We calculate $T_{{\rm c}}$ by solving 
\'Eliashberg's equation (Fig.~\ref{fig:Fey}(a)).
In the equation, the normal self-energy and the effective interaction
are obtained within the third-order perturbation
with respect to $U$(Figs.~\ref{fig:Fey}(b) and ~\ref{fig:Fey}(c)).
The diagrams enclosed by a dashed line in Fig.~\ref{fig:Fey}(c) are
the vertex correction terms which are not direct contributions from spin
fluctuations. The other diagrams are included in RPA and FLEX.
We call the latter 'RPA-like diagrams' in this paper.
To satisfy Luttinger's theorem, 
that is, the conservation law of particle number,
we adjust the chemical potential $\mu$ using the secant method.
To solve ${\rm\acute{E}}$liahberg's equation by using the
power method algorithm, we have to calculate the
summation over the momentum and the frequency space. 
Since all summations are in convolution forms, 
we can carry them out using the algorithm of 
the fast Fourier transformation.
For the frequency, irrespective of the temperature,
we have 1024 Matsubara frequencies. Therefore, we calculate throughout
in the temperature region $T\ge T_{\rm lim}$
, where $T_{\rm lim}$ is the lower limit temperature
 for reliable numerical calculation,
which is estimated to be approximately 
$2.5\times 10^{-3}(>W/(2\pi\times 1024)\simeq 2.2\times 10^{-3})$, 
where $W$ is the bandwidth;
we divide a primitive cell into 128$\times$128 meshes.

The possible five candidates of unconventional pairing symmetry in our two-dimensional 
model are shown in Fig.~\ref{fig:candidate}.
In our calculation,
we obtain the maximum eigen value
$\lambda_{\rm max}^{\rm SS}$ of spin-singlet pairing
belonging to the $d$-wave and the maximum eigen value
$\lambda_{\rm max}^{\rm ST}$
of spin-triplet pairing
belonging to the $f_{y(y^{2}-3x^{2})}$-wave.
The $f_{y(y^{2}-3x^{2})}$-wave pairing state has been proposed 
in several studies of other similar models~\cite{rf:KA,rf:INY,rf:KTA}.
The calculated eigenvalues $\lambda_{\rm max}^{\rm ST}$ and 
$\lambda_{\rm max}^{\rm SS}$ for various values of $T$
are show in Fig.~\ref{fig:Eigen}(a). 
From Fig.~\ref{fig:Eigen}(a), we can see that 
when we estimate $W(\simeq 13t_{3})\simeq 0.5$ eV from band structure 
calculation, 
our $T_{\rm c}$ is estimated to be far below 1 K, 
which is low compared with the observed $T_{\rm c}\simeq 5$K.

To examine how the vertex corrections influence  $T_{\rm c}$,
we calculate 
the eigenvalues $\lambda_{\rm RPA-like}$ 
of ${\rm\acute{E}}$liahberg's equation
by including only RPA-like 
diagrams of anomalous self-energies up to the third order.
We compare $\lambda_{\rm RPA-like}$ 
with $\lambda_{\rm TOPT}$ calculated by including full diagrams of
anomalous self-energies up to third order.
The calculated eigenvalues  $\lambda_{\rm RPA-like}^{f}$  
(for $f$-wave triplet pairing )
and $\lambda_{\rm RPA-like}^{d}$  
(for $d$-wave singlet pairing )
are shown in 
 Fig. ~\ref{fig:Eigen}(b). 
In RPA-like calculation,
the momentum dependence of static bare susceptibility 
$\chi_{0}({\bf q},0)$ is important for pairing symmetry.
The calculated results of 
$\chi_{0}({\bf q},0)$ are shown in Fig.~\ref{fig:X0}.
In this figure, we can observe a broad peak around M the points and 
a sharp peak around the $\Gamma$ point.
From Fig.~\ref{fig:Eigen}(b), we can see that the vertex correction terms
reduce the eigenvalues of spin-singlet pairing, on the other hand
in the case of spin-triplet pairing,
vertex correction terms work in favor
of realizing the spin-triplet superconductivity.
The fact that the vertex correction for a simple one-boson propagation is 
important for spin-triplet pairing, 
 is the same as the case of Sr$_2$RuO$_4$\cite{rf:Sr214}.
As a result, we have 
$\lambda_{\rm max}^{\rm SS}\simeq\lambda_{\rm max}^{\rm ST}$
within our simple model.

We also calculate 
the self-energy 
$\Sigma_{n}^{R}({\bf k},\omega)$ using the Pad${\rm\acute{e}}$ approximation
(whose results are not presented in this paper).
The $\omega$-dependence 
of both ${\rm Re}\Sigma_{n}^{R}({\bf k}_{f},\omega)$
 and ${\rm Im}\Sigma_{n}^{R}({\bf k}_{f},\omega)$
(where ${\bf k}_{f}$ is a Fermi momentum) near $\omega=0$, are respectively
given by ${\rm Re}\Sigma_{n}^{R}({\bf k}_{f},\omega)\propto -\omega$ 
and ${\rm Im}\Sigma_{n}^{R}({\bf k}_{f},\omega)\propto -\omega^{2}$
This behavior is the same as that for a conventional Fermi liquid.
Therefore, the perturbation treatment seems 
to work well in our calculation.

In summary, we have investigated 
the possibility of unconventional superconductivity 
originating from the electron correlation effects,
on the basis of the Hubbard model whose dispersion of the bare energy
band is obtained using the band structure calculation of 
Na$_x$CoO$_2$$\cdot y$H$_2$O.
In the case of spin-singlet pairing, 
we have obtained the maximum eigen value 
$\lambda_{\rm max}^{\rm SS}$
belonging to the $d$-wave.
In the case of spin-triplet pairing, 
we have obtained the maximum eigen value
$\lambda_{\rm max}^{\rm ST}$ 
belonging to the $f_{y(y^{2}-3x^{2})}$-wave.
It is found that the vertex correction terms
reduce eigenvalues of spin-singlet pairing, on the other hand
in the case of spin-triplet pairing,
vertex correction terms work in favor
of realizing the spin-triplet superconductivity.
As a result, it was found that 
$\lambda_{\rm max}^{\rm SS}\simeq\lambda_{\rm max}^{\rm ST}$ and 
the superconducting transition temperatures are estimated to be low 
compared with the observed $T_{\rm c}\simeq 5$K within our simple model.
So we cannot decide spin-singlet pairing or spin-triplet pairing, and 
can not get reasonable transition temperatures 
of unconventional pairing state.
This fact indicates that we have to
 use more a realistic model such as a multi-orbital model or 
the other mechanism such as the phonon-mediated
mechanism may become 
dominant for the superconductivity in real system~\cite{rf:nmr2}.


\section*{Acknowledgments}
One of the authors(Y.N.) 
greatly thanks to Prof. N. Hamada for allowing us to use his 
FLAPW code and acknowledges Mr. T. Imai and Dr. M. Usuda for 
advising on the band structure calculation of Na$_x$CoO$_2$$\cdot
y$H$_2$O. Numerical computation in this work was carried out at the Yukawa
 Institute Computer Facility.

\clearpage
\setcounter{figure}{0}
\begin{figure}
\includegraphics[width=10cm,height=8cm]{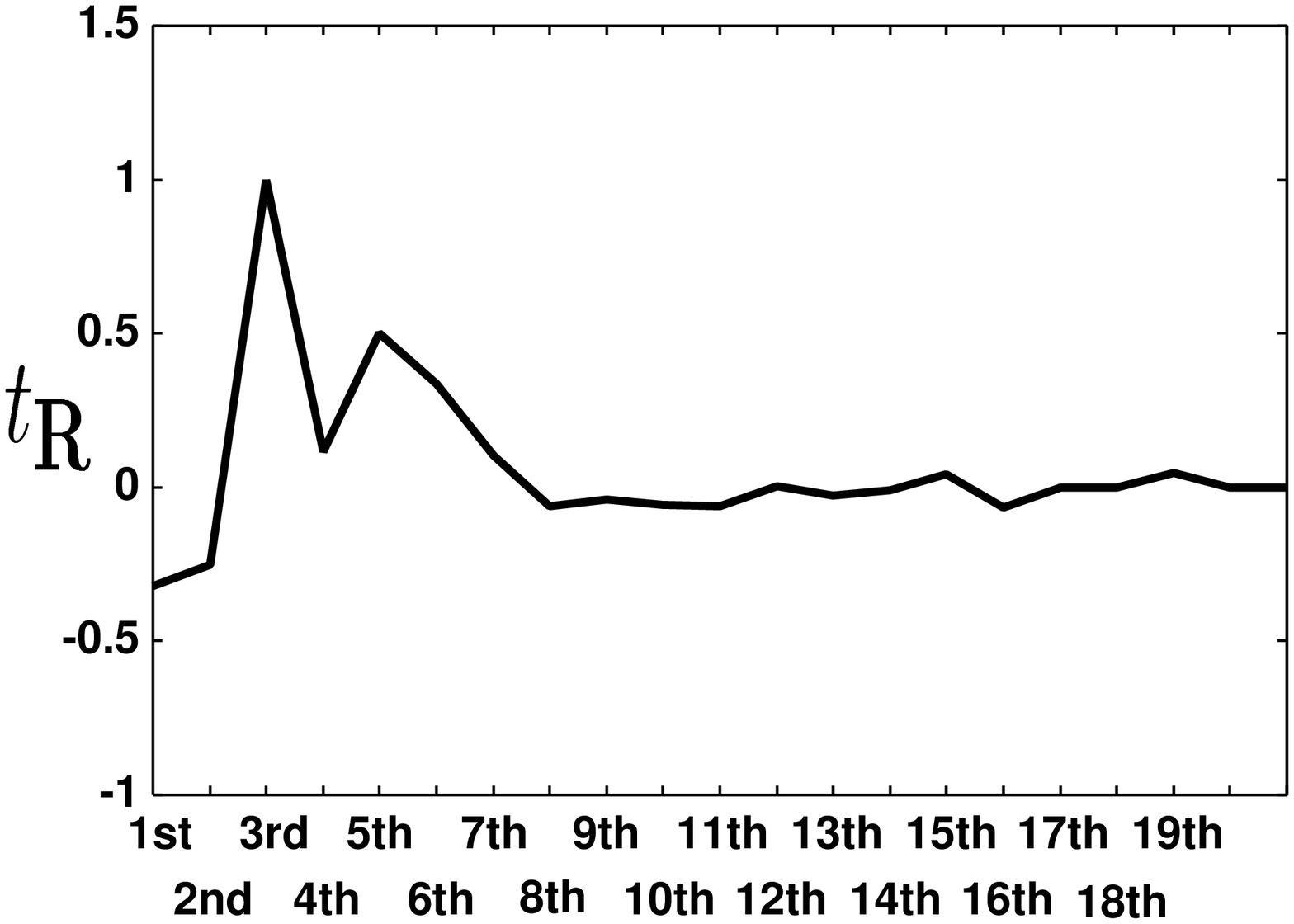}
\caption{The values of the coefficients $t_{\bf R}$}
\label{fig:exp}
\end{figure}
\clearpage

\begin{figure}
\begin{center}
\includegraphics[width=10cm,height=13cm]{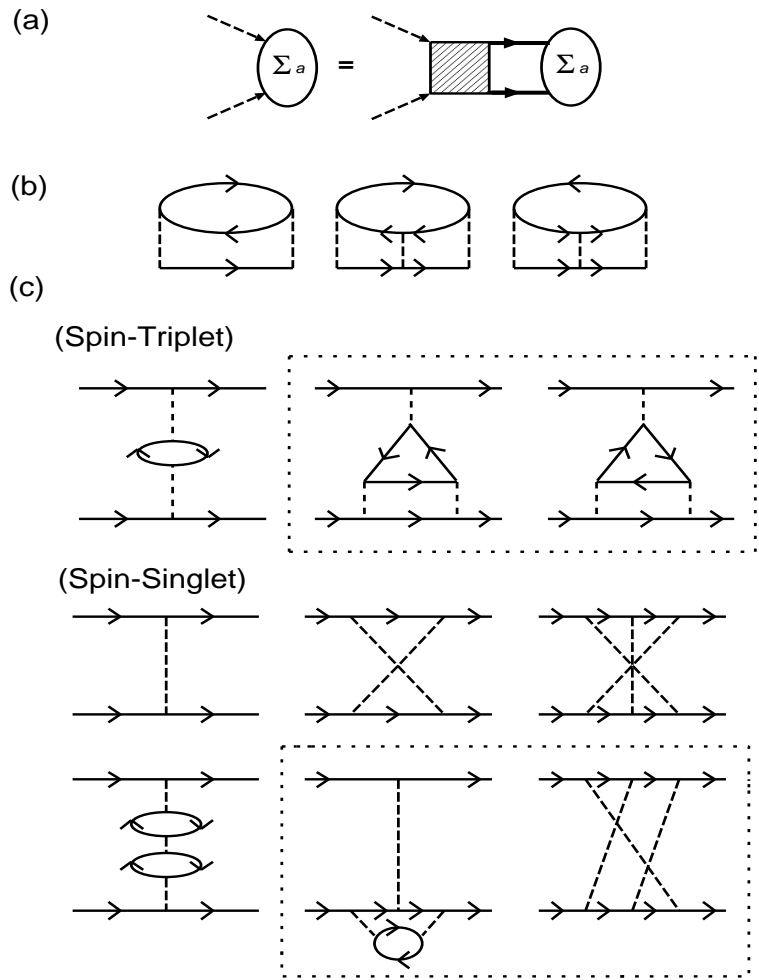}
\end{center}
\caption{(a)\'Eliashberg's equation.
The thick line represents Green's function
 with self-energy correction. The shaded rectangle represents
 the effective interaction.(b) Feynman diagrams of the
 normal self-energy up to the third order.
(c) Feynman diagrams of the effective interaction up to the third
 order. Solid and dashed lines correspond to the bare Green's function
 and the interaction, respectively.
 We omit writing the diagrams given by turning the
vertex correction terms in this figure upside down.}

\label{fig:Fey}
\end{figure}

\clearpage
\begin{figure}
\includegraphics[width=9cm,height=16cm]{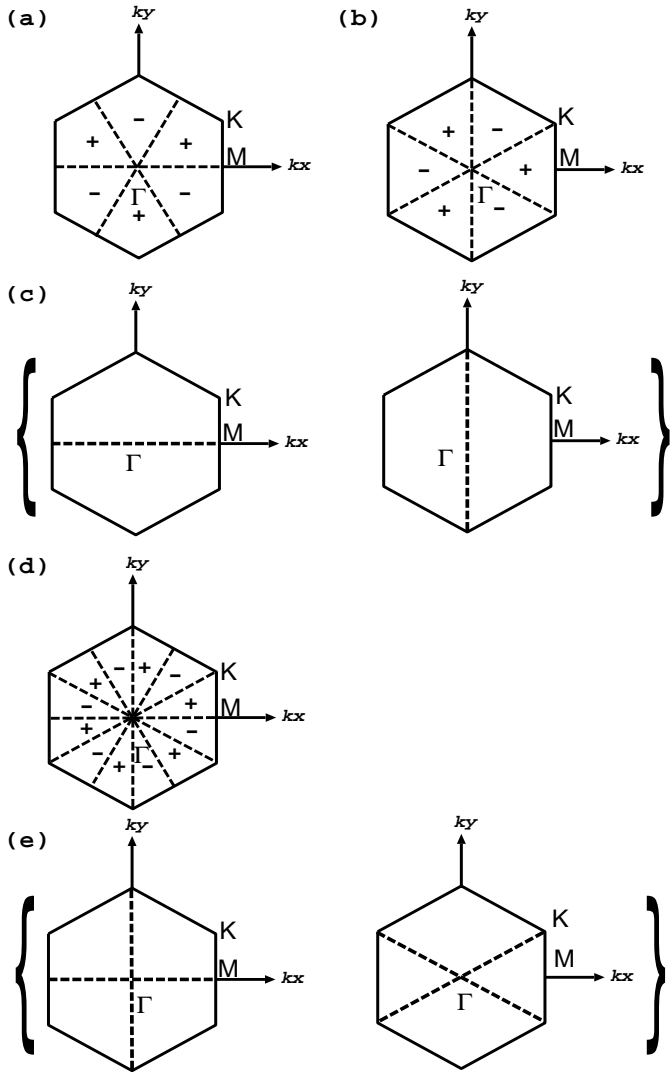}
\caption{The possible five candidates of unconventional
pairing symmetry in our
 model.(a)spin-triplet $f_{y(y^{2}-3x^{2})}$-wave(B$_1$ representation of D$_6$
group), (b)spin-triplet
$f_{x(x^{2}-3y^{2})}$-wave(B$_2$ representation of D$_6$ group),
 (c)spin-triplet $p$-wave(E$_1$ representation of D$_6$ group), (d)spin-singlet
$i_{xy(x^{2}-3y^{2})(y^{2}-3x^{2})}$-wave(A$_2$ representation of D$_6$
 group), (e)spin-singlet $d$-wave(E$_2$ representation of D$_6$ group)}
\label{fig:candidate}
\end{figure}

\clearpage
\begin{figure}
\includegraphics[width=13cm,height=20cm]{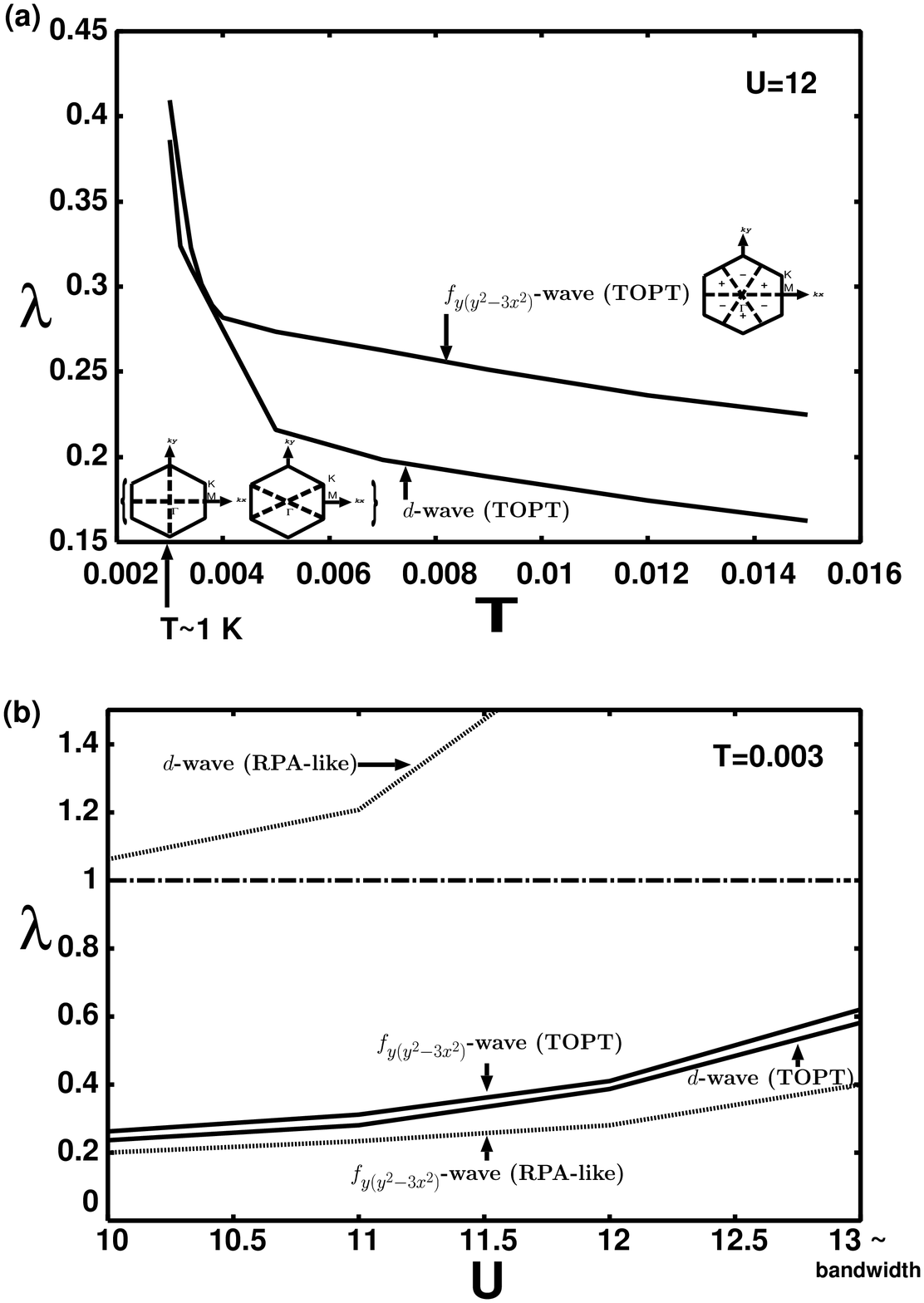}
\caption{(a)The calculated eigenvalues for various values of $T$
(b)The calculated  eigenvalues $\lambda_{\rm TOPT}$ and
$\lambda_{\rm RPA-like}$ for various values of $U$}
\label{fig:Eigen}
\end{figure}

\clearpage
\begin{figure}
\includegraphics[width=12cm,height=14cm]{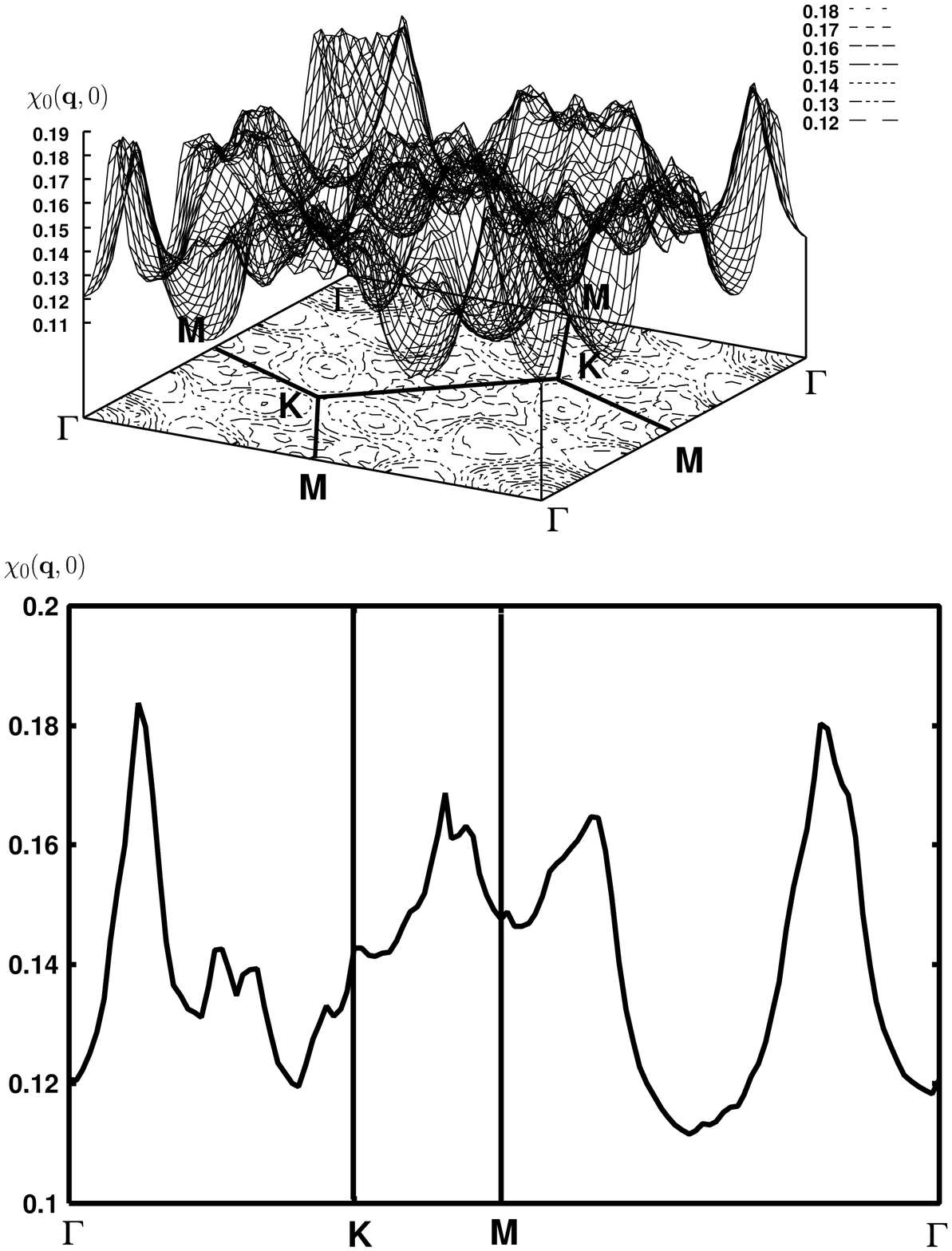}
\caption{The momentum dependence of the static bare susceptibility.}
\label{fig:X0}
\end{figure}

\begin{thebibliography}{99}
\bibitem{rf:Taka}K. Takada {\it et. al}
: Nature {\bf 422} (2003) 53.
\bibitem{rf:z1}H. Sakurai {\it et. al}: Phys. Rev. B {\bf 68} (2003)132507
\bibitem{rf:z2}T. Waki {\it et. al}: cond-mat/0306036.
\bibitem{rf:z3}W.Higemoto {\it et. al}: cond-mat/0310324.
\bibitem{rf:z4}W. Koshibae and S. Maekawa: cond-mat/0306696.
\bibitem{rf:z5}G. Baskaran: Phys. Rev. Lett. {\bf 91}(2003) 097003.
\bibitem{rf:z6}B. Kumar and B.S. Shastry: Phys. Rev. B {\bf 68} (2003) 104508.
\bibitem{rf:z7}Q.-H.Wang {\it et. al}: cond-mat/0304377.
\bibitem{rf:z8}M. Ogata: J. Phys. Soc. Jpn. {\bf 72} (2003) 1839.
\bibitem{rf:z9}A. Tanaka and X.Hu: Phys. Rev. Lett. {\bf 91} (2003) 257006.
\bibitem{rf:z10}Y. Tanaka, Y. Yanase, and M. Ogata: cond-mat/0311266.
\bibitem{rf:heat}G. Cao {\it et. al}: J. Phys.: Condens. Matter {\bf 15} (2003) L519-L525
\bibitem{rf:nmr1}T. Waki {\it et. al}: cond-mat/0306036.
\bibitem{rf:nmr2}Y. Kobayashi {\it et. al}: J. Phys. Soc. Jpn {\bf 72}
	(2003) 2161, J. Phys. Soc. Jpn. {\bf 72} (2003) 2453.  
\bibitem{rf:nmr3}T. Fujimoto {\it et. al}: cond-mat/0307127.
\bibitem{rf:ishi}K. Ishida {\it et. al}: cond-mat/0308506.
\bibitem{rf:INY}H. Ikeda, Y. Nisikawa and K. Yamada:
	J. Phys. Soc. Jpn.{\bf 73} (2004) 17.
\bibitem{rf:KTA}K. Kuroki, Y. Tanaka and R. Arita:cond-mat/0311619 
\bibitem{rf:Da}M. Vojta and E. Dagotto: Phys. Rev. B {\bf 59} (1999) R713.
\bibitem{rf:UsuNisi} Y. Nisikawa and M. Usuda: unpublished.
\bibitem{rf:Shin}D. J. Singh: Phys. Rev. B {\bf 61} (2000) 13397. 
\bibitem{rf:KA}K. Kuroki and R. Arita: Phys. Rev. B {\bf 63} (2001) 174507
\bibitem{rf:Sr214} T. Nomura and K. Yamada: J. Phys. Soc. Jpn.{\bf 69}
	(2000) 3678.
\end{thebibliography}
\end{document}